\documentclass[amsmath,amssymb,superscriptaddress,nobalancelastpage,aps,prb,twocolumn,10pt]{revtex4-1}
\usepackage{graphicx}
\usepackage{epstopdf}
\usepackage{varioref}
\usepackage{xr-hyper}
\usepackage{xcolor}
\definecolor{bl}{rgb}{0.0,0.2,0.6}
\usepackage{hyperref}
\usepackage[english]{babel}

\def\Re{\textrm{Re}}
\def\Im{\textrm{Im}}

\begin{document}

\title{Electron scattering, charge order, and pseudogap physics in La$_{1.6-x}$Nd$_{0.4}$Sr$_x$CuO$_4$:
An angle resolved photoemission spectroscopy study}

\author{C. E. Matt}
\affiliation{Swiss Light Source, Paul Scherrer Institut, CH-5232 Villigen PSI, Switzerland}
\affiliation{Laboratory for Solid State Physics, ETH Z\"{u}rich, CH-8093 Z\"{u}rich, Switzerland}
\author{C. G. Fatuzzo}
\affiliation{Institute for Condensed Matter Physics, \'{E}cole Polytechnique F\'{e}d\'{e}rale
de Lausanne (EPFL), CH-1015 Lausanne, Switzerland}
\author{Y. Sassa}
\affiliation{Swiss Light Source, Paul Scherrer Institut, CH-5232 Villigen PSI, Switzerland}
\affiliation{Laboratory for Solid State Physics, ETH Z\"{u}rich, CH-8093 Z\"{u}rich, Switzerland}
\affiliation{Department of Physics and Astronomy, Uppsala University, S-75121 Uppsala, Sweden}
\author{M.~M\aa nsson }
\affiliation{Institute for Condensed Matter Physics, \'{E}cole Polytechnique F\'{e}d\'{e}rale
	de Lausanne (EPFL), CH-1015 Lausanne, Switzerland}
\affiliation{KTH Royal Institute of Technology, Materials Physics, S-164 40 Kista, Sweden}
\affiliation{Laboratory for Neutron Scattering, Paul Scherrer Institut, CH-5232 Villigen, Switzerland}
\author{S. Fatale}
\affiliation{Institute for Condensed Matter Physics, \'{E}cole Polytechnique F\'{e}d\'{e}rale
	de Lausanne (EPFL), CH-1015 Lausanne, Switzerland}

\author{V. Bitetta}
\affiliation{Institute for Condensed Matter Physics, \'{E}cole Polytechnique F\'{e}d\'{e}rale
	de Lausanne (EPFL), CH-1015 Lausanne, Switzerland}
\author{X. Shi}
\affiliation{Swiss Light Source, Paul Scherrer Institut, CH-5232 Villigen PSI, Switzerland}
\author{S. Pailh\`{e}s}
\affiliation{Laboratory for Neutron Scattering, Paul Scherrer Institut, CH-5232 Villigen, Switzerland}
\affiliation{Institut Lumi\`{e}re Mati\`{e}re, UMR5306 Universit\'{e} Lyon 1-CNRS, Universit\'{e} de Lyon 69622 Villeurbanne}
\author{M. H. Berntsen}
\affiliation{KTH Royal Institute of Technology, Materials Physics, S-164 40 Kista, Sweden}
\author{T.\ Kurosawa}
\affiliation{Department of Physics, Hokkaido University - Sapporo 060-0810, Japan}
\author{M.\ Oda}
\affiliation{Department of Physics, Hokkaido University - Sapporo 060-0810, Japan}
\author{N.\ Momono}
\affiliation{Department of Applied Sciences, Muroran Institute of Technology, Muroran 050-8585, Japan}
\author{O.~J.~Lipscombe}
\affiliation{H.\ H.\ Wills Physics Laboratory, University of Bristol, Bristol, BS8 1TL, United Kingdom}
%----------------------------------------
\author{S.~M.~Hayden}
\affiliation{H.\ H.\ Wills Physics Laboratory, University of Bristol, Bristol, BS8 1TL, United Kingdom}
\author{J. -Q. Yan}
\affiliation{Materials Science and Technology Division, Oak Ridge National Laboratory, Oak Ridge, Tennessee 37831, United States}
\author{J. -S. Zhou} 
\affiliation{Texas Materials Institute, University of Texas at Austin, Austin, Texas 78712, USA} 
\author{J. B. Goodenough}
\affiliation{Texas Materials Institute, University of Texas at Austin, Austin, Texas 78712, USA} 
\author{S.~Pyon}
\affiliation{Department of Advanced Materials, University of Tokyo, Kashiwa 277-8561, Japan}
\author{T. Takayama}
\affiliation{Department of Advanced Materials, University of Tokyo, Kashiwa 277-8561, Japan}
\author{H.~Takagi}
\affiliation{Department of Advanced Materials, University of Tokyo, Kashiwa 277-8561, Japan}
\author{L.~Patthey}
\affiliation{Swiss Light Source, Paul Scherrer Institut, CH-5232 Villigen PSI, Switzerland}
\author{A.~Bendounan}
\affiliation{Swiss Light Source, Paul Scherrer Institut, CH-5232 Villigen PSI, Switzerland}
\author{E.~Razzoli}
\affiliation{Swiss Light Source, Paul Scherrer Institut, CH-5232 Villigen PSI, Switzerland}
\author{M.~Shi}
\affiliation{Swiss Light Source, Paul Scherrer Institut, CH-5232 Villigen PSI, Switzerland}
\author{N.C. Plumb}
\affiliation{Swiss Light Source, Paul Scherrer Institut, CH-5232 Villigen PSI, Switzerland}
\author{M. Radovic}
\affiliation{Swiss Light Source, Paul Scherrer Institut, CH-5232 Villigen PSI, Switzerland}
\author{M. Grioni}
\affiliation{Institute for Condensed Matter Physics, \'{E}cole Polytechnique F\'{e}d\'{e}rale
	de Lausanne (EPFL), CH-1015 Lausanne, Switzerland}
\author{J.~Mesot}
\affiliation{Institute for Condensed Matter Physics, \'{E}cole Polytechnique F\'{e}d\'{e}rale
	de Lausanne (EPFL), CH-1015 Lausanne, Switzerland}
\affiliation{Laboratory for Solid State Physics, ETH Z\"{u}rich, CH-8093 Z\"{u}rich, Switzerland}
\affiliation{Laboratory for Neutron Scattering, Paul Scherrer Institut, CH-5232 Villigen, Switzerland}
\author{O.~Tjernberg}
\affiliation{KTH Royal Institute of Technology, Materials Physics, S-164 40 Kista, Sweden}
\author{J.~Chang}
\affiliation{Swiss Light Source, Paul Scherrer Institut, CH-5232 Villigen PSI, Switzerland}
\affiliation{Institute for Condensed Matter Physics, \'{E}cole Polytechnique F\'{e}d\'{e}rale
	de Lausanne (EPFL), CH-1015 Lausanne, Switzerland}
\affiliation{Physik-Institut, Universit\"{a}t Z\"{u}rich, Winterthurerstrasse 190, CH-8057 Z\"{u}rich, Switzerland}

\begin{abstract}
We report an angle-resolved photoemission study of the charge stripe ordered 
La$_{1.6-x}$Nd$_{0.4}$Sr$_x$CuO$_4$ system.
A comparative and quantitative line shape analysis is presented as the system evolves from 
the overdoped regime into  the charge ordered phase. On the overdoped side ($x=0.20$), a normal state anti-nodal 
spectral gap opens upon cooling below ~80~K. In this process spectral weight is preserved but redistributed to larger energies.
%, but preserves, spectral weight.
A correlation between this spectral gap and electron scattering is found.    
A different lineshape is observed
in the antinodal region of charge ordered Nd-LSCO $x=1/8$. Significant low-energy spectral weight appears 
to be lost. These observations are discussed in terms of spectral weight redistribution and
gapping %of spectral weight 
originating from charge stripe ordering. 
%When charge order is negligibly weak, 
%a correlation between 
%electron scattering and the pseudogap is found. 
\end{abstract}

\maketitle

\section{Introduction}
Partial gapping of spectral weight in absence of any metal instability appears in
many strongly correlated electron systems~\cite{mannellaNAT2005,borisenkoPRL2008,uchidaPRL2011,ChandPRB2012}. This so-called pseudogap phenomenon
is, for example, found in the normal state of charge-density-wave (CDW) systems,  above 
the CDW onset temperature~\cite{borisenkoPRL2009}. A pseudogap phase has also been reported in 
the normal state of high-temperature cuprate superconductors. 
The nature of these pseudogaps is still being debated~\cite{normanAP2005,chatterjeeNATPHYS2010,jleeSCIENCE2009,rdaouNAT2010,mhashimotoNATPHYS2010,tkondoNAT2009,tkondoNATPHYS2011,tkondoPRL2013,skawasakiPRL2010,yheSCIENCE2014}. 
Recently, it has become 
clear that charge ordering is a universal property of hole doped 
cuprates~\cite{twuNAT2011,twuNATCOMM2013,jchangNATPHYS2012,ghiringhelliSCIENCE2012,achkarPRL2012,
	huckerPRB2014,blancocanosaPRB2014,doironleyraudPRX2013,tabisNATCOMM2014,dasilvanetoSCIENCE2014,cominSCIENCE2014,fujitaSCIENCE2014,christensenARXIV2014,thampyPRB2014,croftPRB2014}.
Around the so-called 1/8-doping, the CDW onset temperature appears 
much before the superconducting transition temperature. 
The normal state of cuprates should hence be revisited to 
 identify a single particle gap from CDW order and 
to investigate the spectral gapping in absence of both superconductivity and CDW order.
We therefore present an angle-resolved photoemission spectroscopy (ARPES) study of 
the well-known charge stripe ordered system La$_{1.6-x}$Nd$_{0.4}$Sr$_x$CuO$_4$ (Nd-LSCO),
in which charge and spin orders are coupled~\cite{tranquadaNAT1995,christensenPRL2007}. 
As shown in the phase diagram (Fig.~1), this material has a strongly suppressed superconducting transition temperature, which %, $T_c$, 
allows 
a low temperature study of the normal state.  
We have studied the spectral lineshape evolution as a function 
of momentum, temperature and doping. 
On the overdoped side, Nd-LSCO $p=0.20$, an antinodal spectral gap is observed.
This gap can be closed by either increasing doping to $p=0.24$, increasing temperature 
to $T\sim 80$~K or moving in momentum towards the zone diagonal.
The normal state gap $\Delta$ redistributes spectral weight up to 
$\sim 2.5\Delta$, but the total weight remains conserved.
Analysis of the spectral lineshape suggests a correlation between the
gap amplitude and electron scattering.
In the underdoped regime $p<0.15$, the antinodal lineshape changes.
Compared to the overdoped side of the phase diagram, a significant 
suppression of spectral weight is observed. This effect is 
discussed in terms of quasiparticle decoherence and competing orders. 
In particular, the idea that charge stripe order can contribute to the 
suppression of antinodal spectral weight is discussed.

\begin{figure}[h!]
\begin{center}
\includegraphics[width=0.42\textwidth]{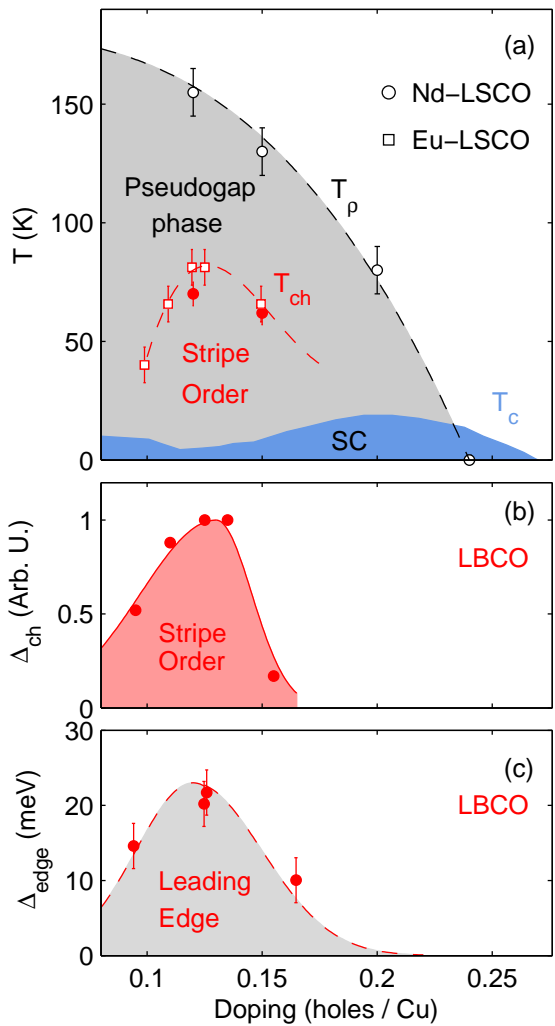}
\end{center}
\caption{(Color online)  (a) Temperature-doping phase diagram of La$_{1.6-x}$Nd$_{0.4}$Sr$_x$CuO$_4$ (Nd-LSCO),
 established by diffraction and resistivity experiments~\cite{rdaouNAT2009,cyrchoinierePHYSC2010,tranquadaNAT1995,swakimotoPRB2003,finkPRB2011}. 
The  temperature scale $T_\rho$ is determined by 
the deviation from  high-temperature linear 
resistivity~\cite{rdaouNAT2009}. The charge ordering temperature ($T_{ch}$) is 
obtained from x-ray  diffraction~\cite{tranquadaNAT1995,swakimotoPRB2003,finkPRB2011}.  
All lines are guides to the eye.
(b) 
Charge stripe order parameter $\Delta_{ch}$, derived from hard x-ray diffraction 
experiments on La$_{2-x}$Ba$_x$CuO$_4$ (LBCO)~\cite{huckerPRB2013}. 
(c) Leading edge gap of LBCO versus doping, from Ref.~\onlinecite{VallaSCIENCE2006}.  }
\label{fig:fig1}
\end{figure}

\begin{figure*}[!]
\begin{center}
\includegraphics[width=0.75\textwidth]{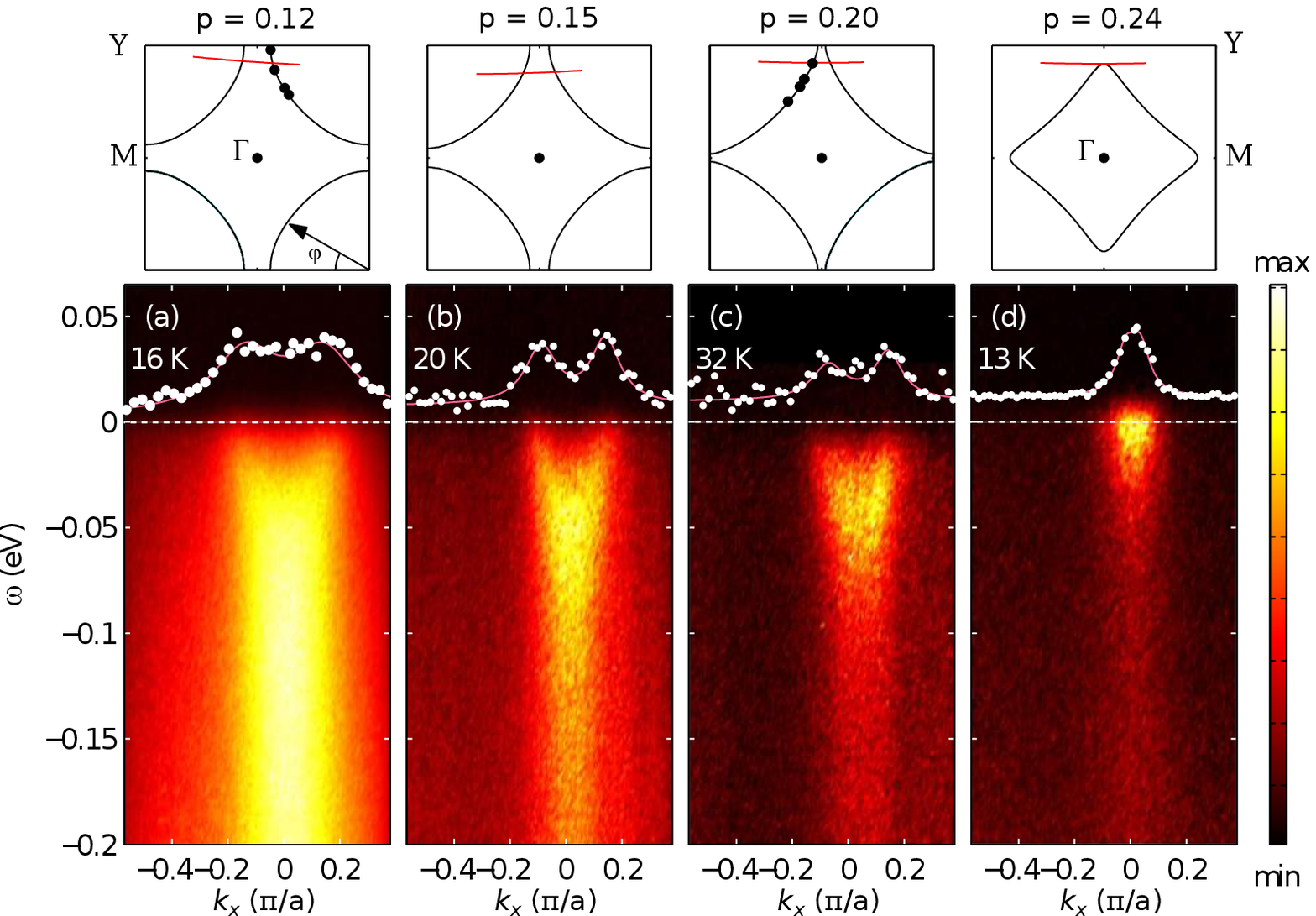}
\end{center}
\caption{(Color online) (a)-(d) Anti-nodal angle-resolved photoemission spectra, taken in the normal state of 
La$_{1.6-x}$Nd$_{0.4}$Sr$_x$CuO$_4$ for different dopings $p=x$ as indicated. Solid white points are momentum disctribution 
curves at the Fermi level, indicated by horizontal dashed lines. Top panels schematically show the Fermi 
surface topology for each of the doping concentrations. 
The red lines indicate the trajectory along which the anti-nodal spectra 
were recorded. Solid black points indicate the underlying Fermi momenta 
at which symmetrized EDCs are shown in Fig. 3(c-d). }
\label{fig:fig2}
\end{figure*}

\section{Methods} 
Our ARPES experiments were carried out at the Swiss Light Source (SLS)
on the Surface and Interface Spectroscopy (SIS) beam line,~\cite{FlechsigAIPCP2004} using 55 eV circular polarized
photons. Single crystals of Nd-LSCO with $x=p=0.12$, 0.15, 0.20 and 0.24 -- grown by the traveling zone
method -- were cleaved \textit{in-situ} under ultra-high vacuum (UHV)  conditions ($\sim0.5 \times 10^{-10}$ mbar)
using a top-post technique or a specially designed cleaving tool~\cite{manssonREVSI2007}. 
Photo-emitted electrons were analyzed using 
a SCIENTA 2002 or a R4000 analyzer. A total energy resolution of $\sim15$ meV was achieved 
with this setup.
Due to matrix element effects, all data were recorded in the second Brillouin zone 
but represented by the equivalent points in the first zone. 
The Fermi level was measured on poly-crystalline copper in thermal and electric contact with the sample. 
Copper spectra were also used to normalize detector efficiencies.\\

\section{Results}
 %\subsection{Electronic structure} 
Normal state ($T\gtrsim T_c$) energy distribution maps  taken in the anti-nodal 
($\pi$,0)-region of Nd-LSCO $x=p=0.12$, 0.15, 0.20, and 0.24
are shown in Fig.~\ref{fig:fig2}. 
As doping $p$ is reduced, the 
"quasiparticle" excitations are gradually broadened.
%spectral weight is progressively shifted to larger 
%energies.
Finite spectral weight at the Fermi level $E_F$ ($\omega=0$) is, however, found for all compositions  even 
deep inside the charge stripe ordered phase~\cite{jchangNEWJP2008}.
It is thus  possible to define the %an 
  underlying  Fermi momenta $k_F$ 
from the maximum intensity of the momentum distribution curves (MDC) at $\omega=0$. 
The Nd-LSCO Fermi surface topology~\cite{ClaessonPRB2009}, shown schematically in Fig.~\ref{fig:fig2},
 is similar to that of La$_{2-x}$Sr$_x$CuO$_4$ (LSCO)~\cite{tyoshidaPRB2006,razzoliNEWJP2010} and Bi2212~\cite{KaminskiPRB06,BenhabibPRL15}.  
A van-Hove singularity crosses $E_F$ at a doping concentration slightly larger than $x = p = 0.20$,
separating electron- from hole-like Fermi surfaces.

\subsection{Spectral lineshapes}
Analysis of symmetrized energy distribution 
curves (EDCs) at $k = k_F$ is a standard method to visualize the existence of 
a spectral gap  near the Fermi level~\cite{normanNAT1998}.
A single-particle gap shifts the spectral weight away 
from the Fermi level and hence produces a double peak structure
in the symmetrized curves. In absence of a spectral gap, 
 the symmetrized EDC at $k_F$ is on the contrary characterized 
 by a lineshape peaked at the Fermi level. %~\cite{Note3}.

\begin{figure*}
 \begin{center}
\includegraphics[width=0.87\textwidth]{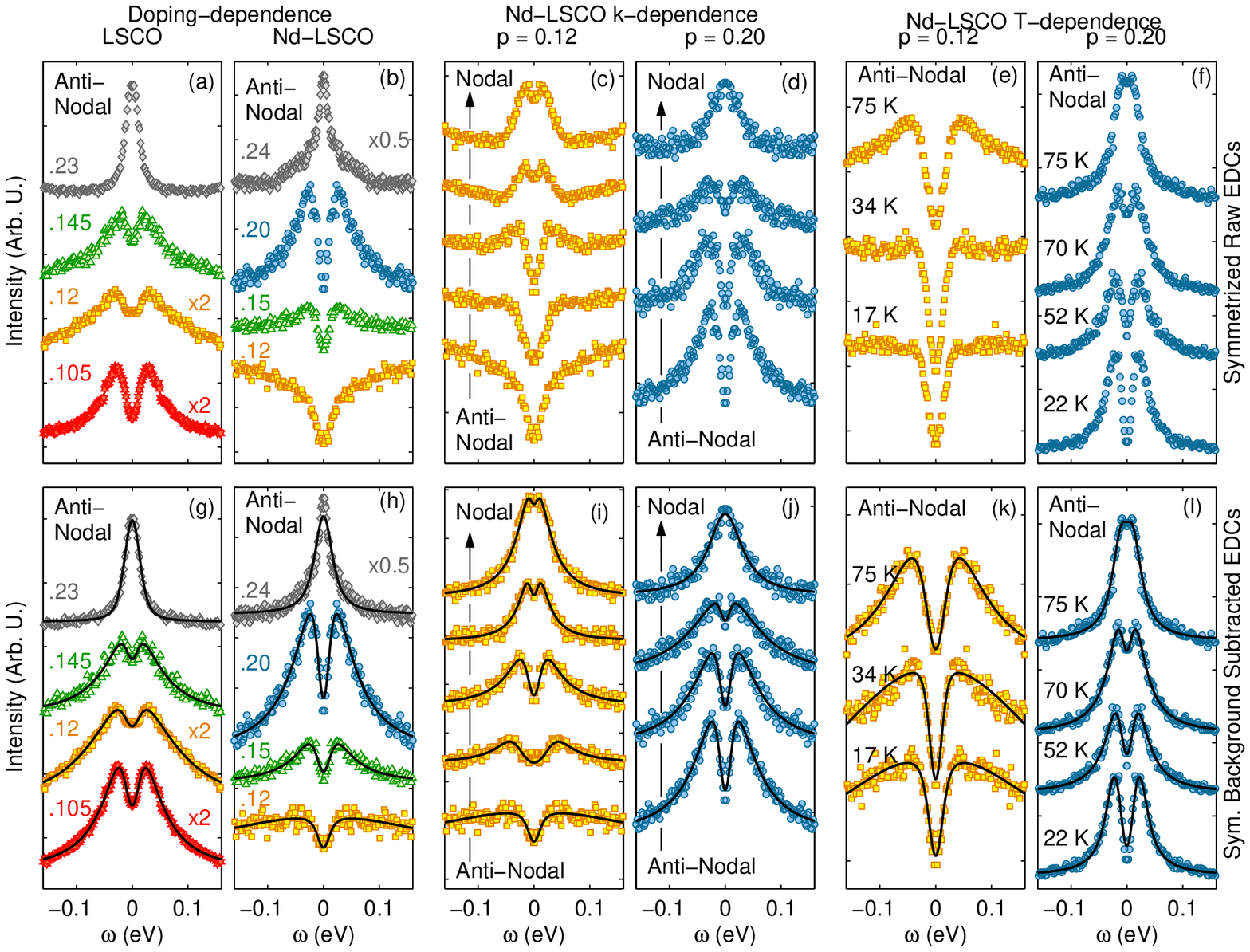}
  \end{center}
  \caption{(Color online) Symmetrized normal state energy distribution curves (EDCs) recorded on La$_{2-x}$Sr$_x$CuO$_4$ (LSCO) and La$_{1.6-x}$Nd$_{0.4}$Sr$_x$CuO$_4$ (Nd-LSCO).
All spectra were taken just above $T_c$. In top panels (a)-(f) are raw symmetrized spectra while in bottom panels (g)-(l) are 
background subtracted spectra. (a-b) Symmetrized EDCs taken in the anti-nodal region, for doping concentrations of LSCO and Nd-LSCO as indicated. ARPES data on LSCO $x=0.105$ and 0.145
were previously presented in Ref.~\onlinecite{mshiPRL2008,mshiEPL2009,ChangPRB2008ARPES} and all LSCO samples were characterized by neutron scattering experiments~\cite{ChangPRL2007,ChangPRB2008,ChangPRB2012}.
(c-d) Momentum dependence of symmetrized energy distribution curves (EDCs) taken at $k_F$ moving from anti-nodal (bottom) to nodal 
(top) region for Nd-LSCO $p=0.12$ and 0.20.  
 (e-f) Temperature dependence of anti-nodal symmetrized EDCs recorded on 
 Nd-LSCO $p=0.12$ and 0.20.
 For clarity, each spectrum has been given an arbitrary vertical shift. Solid lines in bottom panels 
 are fits, see text for an explanation. }
  \label{fig:fig3}
  \end{figure*}

For overdoped LSCO and Nd-LSCO $p \sim 0.24$, the anti-nodal spectra 
have a Voigt-like profile (see top spectrum of Fig.~\ref{fig:fig3}a,b) just above $T_c$, suggesting 
 resolution limited gapless excitations.  
At slightly lower doping in Nd-LSCO $p=0.20$,  a clear  spectral gap $\Delta\sim 25-30$ meV is found in the anti-nodal region for $T \sim T_c$  (Fig.~\ref{fig:fig3}b). 
Similar line-shapes of the ARPES spectra were obtained on 
Nd-LSCO $p\sim0.15$ and LSCO with $p=0.105, 0.12$ and 0.15, see Fig.~\ref{fig:fig3}a,b.
As in Bi2212 and Bi2201~\cite{mhashimotoNATPHYS2014,chatterjeePNAS2011,ktanakaSCIENCE2006}, a dramatic change of anti-nodal line shape appears for underdoped Nd-LSCO (Fig.~\ref{fig:fig3}b.).
The peaked lineshape structure -- found for 
Nd-LSCO $p=0.15$ and 0.20 -- is strongly depleted.

 A similar evolution of the line-shape is found when moving 
 from the anti-nodal to the nodal region in Nd-LSCO at $p=0.12$ (Fig.~\ref{fig:fig3}c). 
 It resembles the doping dependence (Fig.~\ref{fig:fig3}b):
 first the double-peaked structure is recovered and second, upon entering the Fermi arc,
 gapless excitations are found~\cite{jchangNEWJP2008}. 
 For comparison, the momentum dependence 
 of the EDC lineshapes in Nd-LSCO $p=0.20$ is shown in Fig.~\ref{fig:fig3}d. 
 At this doping, a peaked structure is found 
 for all underlying Fermi momenta. 
 (see Fig.~\ref{fig:fig3}d). The  temperature dependence of antinodal spectra 
 are also very different in Nd-LSCO $p=0.12$ and 0.20 -- see Fig.~\ref{fig:fig3}(e,f) and \ref{fig:fig5}.
 For $p=0.20$, the normal state gap closes at $T\approx80$~K, while it persists in 
 the stripe order $p=0.12$ compound. Furthermore, the peaked structure 
 in the symmetrized EDC lineshape becomes more pronounced in $p=0.20$ upon cooling (Fig.~\ref{fig:fig3}f).
 The opposite trend is observed at 0.12 doping. In fact, as in Bi2201~\cite{mhashimotoNATPHYS2014},
 a much sharper anti-nodal line-shape is found at 75~K compared to 17~K.
 Finally, the spectral gap in $p=0.20$ seems to conserve 
 but redistribute the spectral weight (Fig.~\ref{fig:fig5}) as it opens 
 upon cooling. In contrast, for underdoped Nd-LSCO $p=0.12$, spectral weight is
 either lost or redistributed in a non-trivial fashion upon cooling.
 The anti-nodal spectra at the anomalous 1/8 doping are thus 
 behaving very differently from what is found in more overdoped 
 samples of Nd-LSCO. The 1/8 anti-nodal spectra are also very 
 different from what is observed in LSCO at similar doping (Fig.~3).

\subsection{Background subtraction}
The raw spectra, described above, are composed of an intrinsic signal on 
top of an extrinsic background. Importantly,  
the extrinsic background has essentially the same profile for all measured
compounds. It is therefore possible to normalize spectral 
intensities relatively to the extrinsic background - see Appendix. 
Anti-nodal spectra were recorded on several cleaved surfaces 
of Nd-LSCO $p=0.12$ and different ratios between signal and 
extrinsic backgrounds were found. As a consequence, 
slightly different raw anti-nodal line-shapes were extracted.
However, once background is subtracted, consistent lineshapes
were reproduced (shown in the Appendix).  
As shown in Fig.~3(g-l), only the antinodal 
lineshape of Nd-LSCO with $p=0.12$ is significantly influenced 
by the background subtraction. For all other spectra, 
the background subtraction has little impact on the 
overall lineshape. In fact, for Nd-LSCO $p=0.12$
the signal is comparable to the background, whereas for 
compounds with $p>0.15$ the signal-to-background ratio
is much larger (see Fig.~\ref{fig:fig5}). %\\[1.5mm]
Again, this is an indication that the 1/8 anti-nodal 
spectra are anomalous.

\section{Discussion} 

\subsection{Lineshape modelling}
Lets start by discussing the spectra on the overdoped side of 
the phase diagram.
Neglecting  matrix 
element effects, the symmetrized intensity $I(k_F,\omega)$ is given by the 
spectral function~\cite{normanNAT1998} 
\begin{equation}
A(k_F,\omega)\sim -\Im\Sigma / [(\omega-\Re\Sigma)^2+\Im\Sigma^2].
\end{equation}
In absence of a spectral gap, $\Re\Sigma =0$ at $k=k_F$  and the spectral function 
is nothing else than a Lorentzian function, when approximating $\Im\Sigma$ by a constant 
$\Gamma$. If $\Im\Sigma=\Gamma$ is comparable to 
the applied energy resolution, a Voigt lineshape is effectively observed. 
This is the case for anti-nodal spectra of 
Nd-LSCO $p=0.24$ (Fig.~\ref{fig:fig3}h). The intrinsic linewidth $\Gamma$ is a measure of the 
"quasiparticle" scattering. With increasing scattering, the linewidth
broadens ($\Gamma$ increases) and the peak amplitude -- sometimes referred to as
the "quasiparticle residue $Z$" -- is lowered. 
In this fashion, a metal can loose its coherence.

In presence of a spectral gap, Eliashberg theory 
applied to the normal state finds 
the Green's function $G(k_F,\omega)=[(\omega+i\Gamma)-\Delta^2/(\omega+i\Gamma)]^{-1}$ 
to be given by two parameters: the gap $\Delta$ and the scattering rate $\Gamma$~\cite{ChubukovPRB2007}. 
This functional form mimics roughly the
observed lineshape, but does not provide a fulfilling description of 
the experimental spectra.   
We, therefore, adopted a simpler phenomenological Green's function, 
$G(k_F,\omega)=[(\omega+i\Gamma)-\Delta^2/\omega]^{-1}$, that contains the same two parameters
and has previously been used to analyze symmetrized energy distribution 
curves~\cite{AKanigelNATPHYS2006,jleeSCIENCE2009,normanPRB1998,FranzPRB1998,mshiPRL2008,mshiEPL2009}. 
The spectral function $A(k_F,\omega)=\pi^{-1}\Im G(k_F,\omega)$ can now be expressed by two dimensionless quantities,  
\begin{equation}
A(x)\sim \frac{1}{\Delta}\frac{\gamma}{(x-1/x)^2+\gamma^2} 
\end{equation}
%now expressed by two dimensionless quantities
where $x=\omega/\Delta$ and $\gamma=\Gamma/\Delta$.
This phenomenological spectral function preserves the Lorentzian lineshape and total spectral weight, but shifts the 
peaks to $x=\pm 1$ ($\omega=\pm\Delta$) while the linewidth $\Gamma/\Delta$ 
is renormalized by the spectral gap.
For a fixed gap $\Delta$, increasing quasiparticle scattering  still  leads
 to a broader line and weaker peak amplitude. 
Absence of a peaked structure may therefore be a signature of
strong quasiparticle scattering.

  \begin{figure}
  	\begin{center}
  		\includegraphics[width=0.45\textwidth]{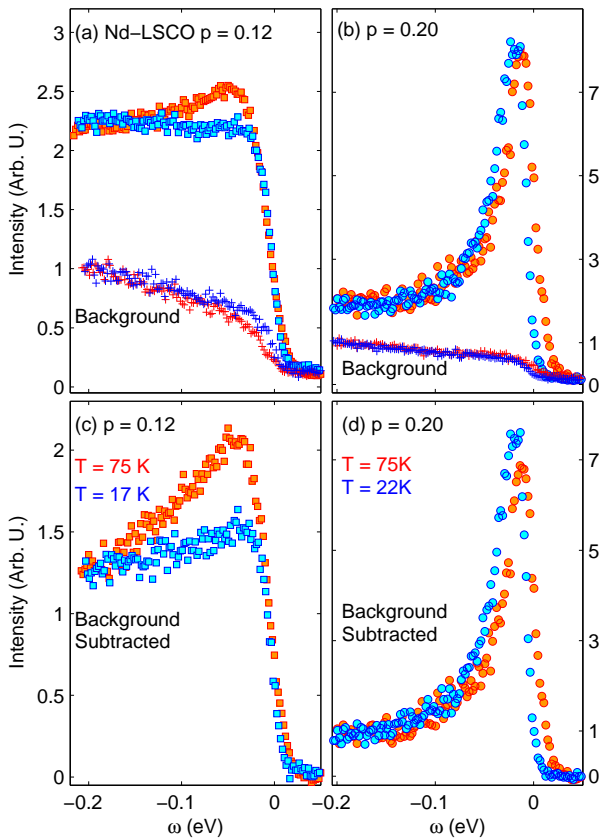}
  	\end{center}
  	\caption{(Color online) Comparison of anti-nodal spectra at $T\sim 20$~K (blue) and $75$~K (red).
  		(a) and (b) show raw energy distribution curves recorded at $k_F$ on Nd-LSCO $p=0.12$ and 0.20
  		with the respective background intensities, measured at momenta far from $k_F$.
  		In (c) and (d), the respective background subtracted curves are compared. } 
  	\label{fig:fig5}
  \end{figure}

  \subsection{Spectral gap and scattering}
\begin{figure*}
 	\begin{center}
 		\includegraphics[width=0.65\textwidth]{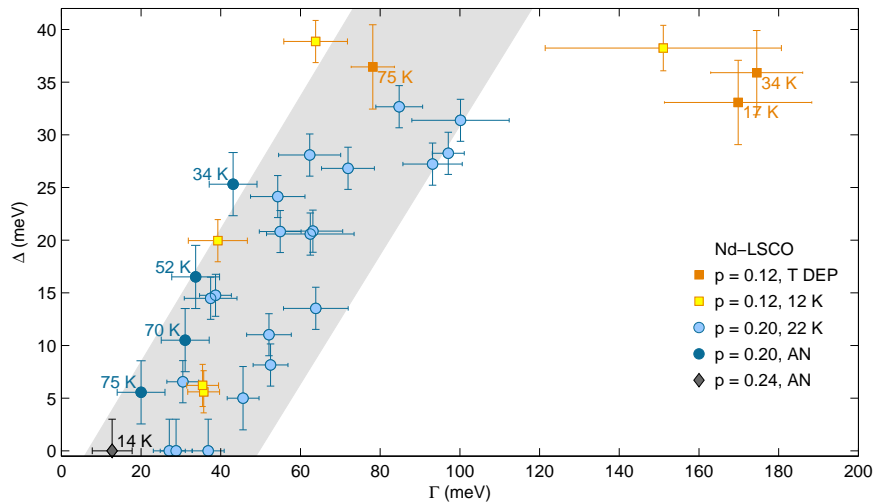}
 	\end{center}
 	\caption{(Color online) Normal state gap $\Delta$ versus the scattering rate $\Gamma$. Both quantities were extracted by fitting 
 		background subtracted	symmetrized energy distribution curves along the underlying Fermi surface of Nd-LSCO $p=0.12$, 0.20 and 0.24, as well as antinodal 
 		spectra versus temperature. The fitting 
 		procedure is explained in the text. Gray shaded area indicates schematically the correlation between 
 		the normal state gap and the electron scattering.} 
 	\label{fig:fig6}
 \end{figure*}

Using Eq.~2, analysis of background subtracted spectra~\cite{GweonPRL2011,FatuzzoPRB2014} was carried out. %For overdoped Nd-LSCO  
Resolution effects 
are modelled by Gaussian convolution 
of the model function $A(k_F,\omega)$ (Eq.~1 and 2).
In this fashion, $\Gamma$  and $\Delta$
were extracted along the underlying Fermi  surface
of Nd-LSCO $p=0.20$. 
As shown  in Fig.~\ref{fig:fig6}, a correlation 
between the gap  $\Delta$ and the scattering rate $\Gamma$ is found.
A similar trend is observed when the gap $\Delta$ is weakened 
by increasing temperature in Nd-LSCO $p=0.20$. 
This relation between the antinodal gap (usually referred to as the pseudogap) and electron scattering is 
consistent with previous observations.
It  is, for example, established 
that the pseudogap is largest near the zone boundary~\cite{normanNAT1998,chatterjeeNATPHYS2010,tkondoNAT2009}. 
At the same time, the scattering rate $\Gamma$ has been shown to increase
when moving from nodal to antinodal regions~\cite{VallaPRL2000,ChangNatCom2013}. 
Furthermore, the photoemission lineshape broadens and the pseudogap increases when doping is reduced 
from the overdoped side of the phase diagram~\cite{chatterjeePNAS2011}. The same trend has 
been reported by STM studies of the density-of-states~\cite{Alldredge2008,KatoJPSJ2008}.
The exact experimental relation between scattering and pseudogap (normal state gap) has, however, 
not been discussed much~\cite{KaminskiPRB2005}. A correlation between scattering and the spectral gap has previously been 
predicted by  dynamical mean-field theory (DMFT) calculations for the Hubbard model~\cite{SenechalPRL2004}. 
Within the DMFT approach~\cite{GullPRL2013,SordiPRB2013,FerreroEPL2009,Alloul2014}, the pseudogap emerges
from electron correlations as a primary effect that, in turn, enhances 
the tendency for the system to undergo superconducting and 
charge-density-wave instabilities, at lower temperatures. Notice however that, as opposed to superconductivity, charge order has not yet been found directly in DMFT calculations. 

 From a different point of view, 
 the pseudogap (normal state gap) emerges as a precursor to superconductivity~\cite{chatterjeeNATPHYS2010,jleeSCIENCE2009,PLeePRX2014}, or 
 as a precursor to an order competing with superconductivity~\cite{cominSCIENCE2014,VishikPNAS2012,EGMoonPRB2010,mhashimotoNATPHYS2014,MHashimotoNMAT2015}. 
 In Bi2201, for example, the charge ordering onset temperature is comparable to the pseudogap temperature scale $T^*$~\cite{cominSCIENCE2014}.
Furthermore, a connection between the charge ordering vector and the  vector 
nesting the Fermi arc tips was found~\cite{cominSCIENCE2014}. It is therefore a possibility that the  
pseudogap is related to fluctuating CDW order. In two-dimensional CDW systems, spectral gaps
are indeed observed above the CDW onset temperature~\cite{MonneyPRB2012,InosovPRB2009}. 
In  cuprates, however, the single particle gap originating from CDW order has 
not been clearly elucidated by ARPES experiments. %\\[1.5mm]

  \subsection{Spectra gaps at 1/8 doping}
 
  It is therefore interesting to discuss the spectral lineshapes at the 1/8-doping, 
  where the charge order parameter has its maximum (Fig. 1). Charge order -- in principle -- should 
  open a single-particle gap somewhere on the Fermi surface~\cite{kissNP2007,ChatterjeeNATCOM2015}.
 It is commonly assumed that the 
stripe ordered ground state found in Nd-LSCO 
is identical to that of La$_{2-x}$Ba$_x$CuO$_4$ (LBCO) and La$_{1.8-x}$Eu$_{0.2}$Sr$_x$CuO$_4$ (Eu-LSCO)
with $p=x\simeq 1/8$~\cite{VojtaAP2009}. All three systems have the same 
low-temperature tetragonal crystal structure, similar thermopower~\cite{JchangPRL2010,LiPRL2007}, and the 
same spin/charge stripe structure~\cite{WilkinsPRB2011,NachumiPRB1998,TranquadaPRB1997,FujitaPRB2004}.
At the particular 1/8 doping -- due to phase competition -- charge stripe order suppresses 
almost completely superconductivity. % due to a competing interaction.
ARPES studies on these stripe ordered systems commonly report 
anti-nodal spectra with little low-energy spectral weight~\cite{VallaSCIENCE2006,rhheNATPHYS2009,ZabalotnyyEPL2009,jchangNEWJP2008,VallaPhysicaC2012}.
Different interpretations have been put forward~\cite{VallaSCIENCE2006,rhheNATPHYS2009}. 
In LBCO it was suggested that the pseudogap (normal state gap) has $d$-wave character and
that the gap amplitude $\Delta$ is maximized at 1/8-doping~\cite{VallaSCIENCE2006} (this result is reproduced in Fig.~1c).
Subsequent experiments reported a correction to the $d$-wave 
symmetry~\cite{rhheNATPHYS2009}. This led to the proposal of a two-gap scenario~\cite{KondoPRL2007,HuefnerRPP2008,MaPRL2008}, 
with an additional spectral gap (of unknown origin)
in the anti-nodal region~\cite{rhheNATPHYS2009}. % was proposed.

In Nd-LSCO $p=0.12$, Fermi arcs with finite length were 
found even at the lowest measured temperatures~\cite{jchangNEWJP2008}.
To access the intrinsic spectral evolution as a function of 
momentum in Nd-LSCO $p=0.12$, background subtracted data
should be considered. In Fig.~\ref{fig:fig3}(i), spectra near the anti-nodal region and 
 close to the tip of the Fermi arc are compared.
 Near to the tip, the spectrum resembles that observed 
 in overdoped Nd-LSCO. Fitting to Eq.~(2) yields  $\Delta=20\pm2$~meV and a scattering constant
 $\Gamma=39\pm8$~meV. This is consistent with the approximate 
 constant ratio of $\Delta/\Gamma$ (see Fig.~\ref{fig:fig6}) found for 
 Nd-LSCO $p=0.20$.
 The lineshape of the anti-nodal spectra is, however,  
 dramatically modified. A similar evolution was found in 
 LBCO~\cite{rhheNATPHYS2009}. It seems that the system 
 has lost coherence.  
 Fitting using Eq.~2, indeed yields much smaller ratios of $\Delta/\Gamma$ -- see Fig.~\ref{fig:fig6}.  
 A sudden quasiparticle decoherence effect is therefore one possible 
 explanation for the different anti-nodal lineshape observed in the underdoped 
 regime.

  \subsection{Effects of competing orders}
Next, we discuss the possible influence of static long-range charge density-wave order. 
For conventional CDW systems, 
the order parameter is identical to the single-particle 
gap~\cite{GruenerRMP1988}, and $\Delta_{ch}$ scales with the lattice distortion 
$u$~\cite{GruenerRMP1988}. By measuring this distortion using hard x-ray diffraction, it was found that 
$\Delta_{ch}$ has a strong doping dependence~\cite{huckerPRB2013} (reproduced in Fig.~\ref{fig:fig1}b) 
-- peaking sharply at the 1/8-doping. 
~Just a slight increase of doping, to say $p=0.15$, 
results in  a single-particle gap $\Delta_{ch}$
renormalized by a factor of five~\cite{huckerPRB2013} (compared to 1/8-doping).
Notice that the charge stripe onset temperature $T_{ch}$ -- observed by x-ray diffraction --
varies more smoothly with doping. Hence, the coupling constant $\alpha=\Delta_{ch}/k_BT_{ch}$
has a strong doping dependence -- being largest at 1/8 doping. It is also around this doping 
that quantum oscillation~\cite{DoironLeyraudNAT2007,SESebastianRPP2012,VignolleCRP2011,barisicNP2013} and transport~\cite{LaliberteNATCOMM2011,LeboeufPRB2011,JchangPRL2010,doironleyraudPRX2013} experiments have revealed the Fermi surface reconstruction
in YBCO and Hg1201. Charge ordering has been proposed as the mechanism responsible for this reconstruction~\cite{LaliberteNATCOMM2011,tabisNATCOMM2014}. 
Strongly coupled charge order is therefore not necessarily in contradiction with the observation 
of quasiparticles with light masses. Interestingly, neither 
the Fermi surface reconstruction nor the effect of charge order have 
been convincingly probed by photoemission spectroscopy.

The observation of an electronic Fermi surface reconstruction is complicated by orthorhombic distortions, that 
fold the bands similarly to what is expected from density-wave orders~\cite{HeNJP2011,MengNAT2009,KingPRL2011}.
Moreover, identification of charge density wave order effects on 
the antinodal lineshape in very underdoped compounds is complicated by superconductivity, pseudogaps and possibly also 
spin-freezing phenomena~\cite{fujitaPRB2002,WakimotoPRB2000}. The choice of Nd-LSCO ensures, due to it's low $T_c$, that superconductivity 
is not influencing the problem. 
Furthermore, in this system spin and charge density wave orderings are coupled~\cite{tranquadaNAT1995},
and hence part of the same phenomenon.

When a spectral gap $\Delta$ opens, low-energy spectral weight is either suppressed or
redistributed in $(k,\omega)$-space. It has, for example, been shown that in Bi2212, pronounced 
redistribution of spectral weight -- extending beyond 200 meV -- appears inside the pseudogap~\cite{MHashimotoNMAT2015}. 
In Fig.~\ref{fig:fig5}b, antinodal spectra
of Nd-LSCO $p=0.20$ display how the normal state gap opens upon cooling. As
the gap opens, spectral weight is transferred to larger energies, while 
the total amount of spectral weight remains approximately constant. This 
rearrangement of spectral weight manifests itself within an energy scale 
$(2-3)\Delta<100$~meV. In the anti-nodal regime of stripe ordered Nd-LSCO $p=0.12$, within 
the same temperature and energy window,
the behaviour is very different (see Fig.~\ref{fig:fig5}a).    
Upon cooling, low-energy ($\omega<100$~meV) spectral weight is removed with 
an apparent net loss of total weight.  
The $k-$dependence in Fig.~\ref{fig:fig3}(c,i), does
not suggest any pile up of spectral weight at other locations in momentum space.
Thus either spectral weight is transferred to  $\omega>5\Delta$, or 
it is simply not conserved. 
A system that undergoes a phase transition may not display 
spectral weight conservation. 
Appearance of charge stripe order in the low-temperature tetragonal 
crystal structure may therefore lead to effective loss of 
spectral weight. 
In that case, stripe order seems to influence mainly the anti-nodal region
and, remarkably, suppression of spectral weight extends up to energies 
as large as 100~meV.

\section{Conclusions} 
In summary, we have presented a systematic 
angle resolved photoemission spectroscopy, normal state study of the  charge stripe ordered cuprate compound La$_{1.6-x}$Nd$_{0.4}$Sr$_x$CuO$_4$ (Nd-LSCO).
By varying the doping concentration, antinodal spectra were recorded 
from the overdoped metallic phase to the 1/8-doping --
where static charge stripe order is stabilized. 
The metallic phase is characterized by gapless excitations even in the 
antinodal region. At $x=0.20$,  a spectral 
gap $\Delta\approx 30$~meV opens in the antinodal region but spectral weight 
remains conserved, although shifted to slightly larger energies.  
Analysis of the line shape suggests a 
correlation between electron scattering and the gap amplitude. 
Finally, for underdoped compounds the anti-nodal lineshape is quite 
different. Upon cooling into the stripe ordered phase, 
spectral weight appears to be lost. 
An additional source for spectral weight suppression 
is therefore proposed, and charge stripe order is discussed 
as an underlying mechanism.\\

\textit{Acknowlegdements.--} This work was supported by the Swiss National Science Foundation 
(through grant Nr 200020-105151, 200021-137783 and its NCCR - MaNEP and Sinergia network Mott Physics Beyond the Heisenberg (HPBH) 
model), the Ministry
 of Education and Science of Japan, and the Swedish Research Council.
Work at ORNL was supported by US-DOE, BES, Materials Sciences and Engineering Division. 
JSZ and JBG were supported by the US NSF (DMR 1122603).
  The photoemission experiments were performed at SLS of the Paul Scherrer Institut,
  Villigen PSI, Switzerland. We thank the X09LA beamline~\cite{FlechsigAIPCP2004} staff
   and Xiaoping Wang for technical support. We wish to thank Nicolas Doiron-Leyraud, Paul Freemann, Markus H\"{u}cker,  Claude Monney, Henrik R\o{}nnow, Louis Taillefer 
and Andr\'{e}-Marie Tremblay for enlightening 
discussions. %\\%[1.0mm]
\clearpage
\section{Appendix A}
\begin{figure}%[!h]
	\begin{center}
		\includegraphics[width=0.35\textwidth]{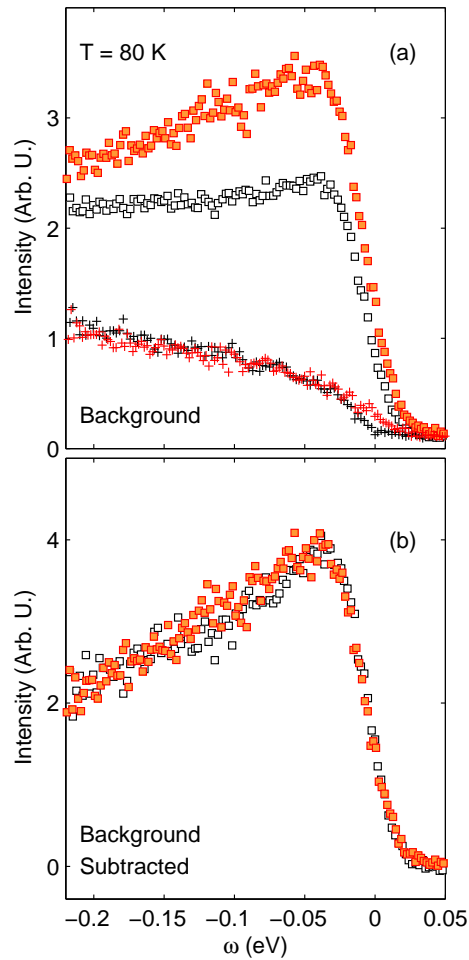} 
	\end{center}
	\caption{(Color online) Comparison of anti-nodal spectra recorded on different surfaces of Nd-LSCO $p=0.12$ at $T=80$~K. 
		(a) Raw spectra at $k_F$ and at momentum $k_{BG}$, representing the extrinsic background. Intensities 
		have been normalized so that the background intensities match across different experiments.
		In this fashion, it shown how the same spectral lineshape can appear different due to 
		a different signal-to-background ratio. Spectra, at $T\sim 80$~K, were taken after 
		cleaving at $T=20$~K (black) and at $80$~K (red).
		(b) Background subtracted spectra, scaled by an arbitrary constant. } 
	\label{fig:fig4}
\end{figure}

All measured ARPES spectra contain background that typically vary slowly with momentum 
and excitation energy $\omega$. The background can be evaluated at momenta far away from 
$k_F$. We found that across all dopings studied, the background has a very similar 
intensity profile as a function of $\omega$. It is thus possible to scale ARPES intensities 
using this background. In Fig.~\ref{fig:fig4}, the background of two  Nd-LSCO $p=0.12$ 
anti-nodal spectra recorded under comparable conditions but on different surfaces.
The background can be scaled / normalized to give an essentially perfect match.
Energy-distribution curves recorded at $k_F$ are, however, displaying different 
intensities and lineshapes. This demonstrates that from experiment  to experiment, 
different signal-to-background ratios are observed. We stress that this effect 
is most visible at $p=0.12$, where anti-nodal spectral weight appears strongly 
suppressed or redistributed. Once the background intensities are
subtracted, the intrinsic lineshape is essentially identical, irrespectively 
of the signal-to-background ratio - see Fig.~\ref{fig:fig4}b. Throughout this 
work, detailed analysis of lineshapes were carried out on the 
background-subtracted data.

%

%%\bibliography{mattetalBIBv2}
%%\bibliographystyle{apsrev4-1}

\end{document}